**The quantum scattering time and its implications on scattering sources in graphene**

X. Hong, K. Zou and J. Zhu

Department of Physics, The Pennsylvania State University, University Park, PA 16802

**Online Supplementary Information Content:**

1. Sample preparation
2. Background subtraction of Shubnikov-de Haas (SdH) oscillations
3. The effect of density inhomogeneity on the quantum scattering time $\tau_q$
4. Determine the concentration of charged impurity $n_{imp}$ at a distance $z$
5. Scattering from charges in the bulk of the $SiO_2$ substrate

## 1. Sample preparation

Kish graphite (Toshiba Ceramics) is used to fabricate Sample A while samples B-D are exfoliated from HOPG graphite (Grade ZYA, GE). Ashing or UV/ozone is used to remove photoresist residue on some $SiO_2$ substrates following standard photolithography used to fabricate alignment marks. Prior to the exfoliation of graphene, all $SiO_2$ substrates are sonicated in acetone for 2 minutes followed by an IPA rinse and drying in a stream of dry $N_2$ gas. Sample A is exfoliated immediately after the cleaning steps. Substrates used for samples B-D are baked at 200°C (B and C) and 50°C (D) in $N_2$ flow for 2 (B and C) and 4 (D) hours before the exfoliation. Subsequent processing steps use standard e-beam lithography and are nominally identical for all samples. Raman spectra taken on devices prepared concurrently with samples A-D resemble that of pristine graphene with no visible D peak.

We do not observe a clear trend between the various preparation recipes and sample mobility, $\tau_q$ or impurity-graphene distance $z$. We, therefore, speculate that uncontrolled spatial variation of $SiO_2$ surface properties, as well as sample preparation conditions (e. g. humidity) may have been the primary reasons behind the observed variations among samples, although variations in sample preparation cannot be ruled out.

## 2. Background subtraction of Shubnikov-de Haas (SdH) oscillations

SdH oscillations in some samples show a slowly varying background, an example of which is given in Figure S1(a). We determine this background by averaging the two dashed curves shown in Fig. S1(a), which are obtained by connecting the maxima and minima in $\rho_{xx}(B)$ respectively. Figure S1(b) shows the same data trace after subtracting the background, from which we determine $\tau_q$ following procedures described in the main text.

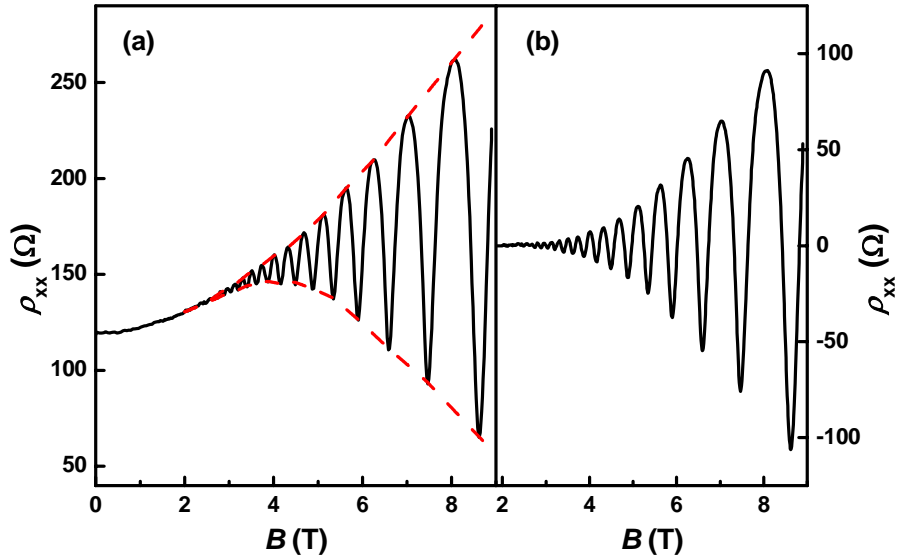

Fig. S1 (a) SdH oscillations of sample D at $n = 5.42 \times 10^{12}/cm^2$ and $T = 10$ K. (b) The same set of oscillations after background subtraction.

## 3. The effect of density inhomogeneity on the quantum scattering time $\tau_q$

In a 2DEG with spatially varying density, the SdH oscillations become an ensemble average of the distributed density profile. Although the resulting oscillations may correspond to a single density and appear well-behaved otherwise [1, 2], the phase shift among different densities effectively reduces the original amplitude of the oscillations, leading to the appearance of a smaller $\tau_q$ with an intercept $\delta\rho_{xx}/\gamma_{th}\rho_0$ larger than the theoretical value of 4 in the Dingle plot, as shown in the inset of Fig. S2. This effect is more pronounced in high-density 2DEGs [1, 2]. Modeling the density with a Gaussian distribution $n\pm\delta n$, we have followed the procedures described in Ref. [2] to simulate the averaged $\rho_{xx}$ and compared with our data. Each individual trace used in the average satisfies $\delta\rho_{xx}/\gamma_{th}\rho_0 = 4$ in the high field limit ($1/B = 0$). The parameters $\tau_q^0$ and $\delta n$ are chosen such that the averaged $\rho_{xx}$ fits the experimental data. An example of this procedure is shown in Fig. S2, where the measured $\delta\rho_{xx}$ is shown in open circles and the solid line represent the simulated $\delta\rho_{xx}$ with $\tau_q^0 = 40$ fs and $\delta n/n = 1.8\%$. The agreement between the two is excellent. The same data yields $\tau_q = 34$ fs with $\delta n = 0$ in Fig. 2(b).

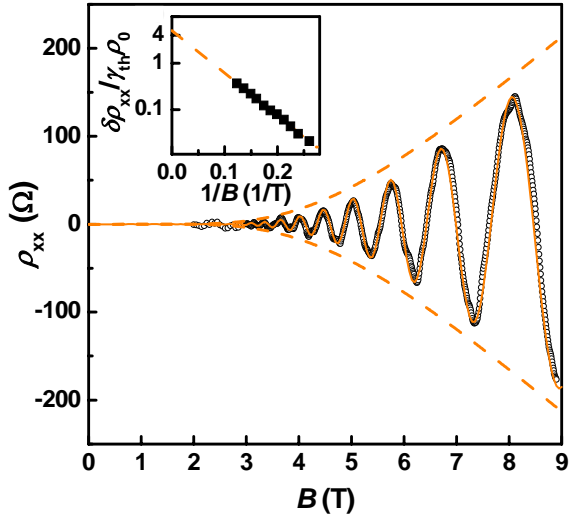

Fig. S2 $\delta\rho_{xx}$ of $n = 3.89\times10^{12}/cm^2$ in sample A (open symbols). Simulated $\delta\rho_{xx}$ with $\tau_q^0 = 40$ fs and $\delta n = 7\times10^{10}/cm^2$ are shown in solid line. The dashed lines indicate $\delta\rho_{xx}$ corresponding to $\tau_q^0 = 40$ fs and $\delta n = 0$. Inset: $\delta\rho_{xx}/\gamma_{th}\rho_0$ vs $1/B$ (the Dingle plot) of data in the main figure showing an intercept of 5 at $1/B = 0$.

Using this method, we have re-examined all the $\tau_q$ fittings and estimated $\delta n$ to be approximately $7\times10^{10}/cm^2$ in sample A and $9\times10^{10}/cm^2$ in sample B in the density range studied. The results are shown in Figs. S3 (a) and (b). $\delta n$ in samples C and D are smaller than the accuracy of our method ($1$-$2\times10^{10}/cm^2$ estimated from fittings similar to Fig. S2) and are therefore neglected. The above correction leads to ~20% increase of $\tau_q$ in sample A and 50% in sample B. As shown in Fig. S3(c), the $\tau_t^{long}/\tau_q^{long}(n)$ after correction (open symbols) in sample A still corresponds to charged impurity located at $z = 0$. The corrected ratio in sample B, on the other hand, now corresponds to charges located at $z = 1$ nm instead of previously calculated 2 nm (Fig. 3(b)).

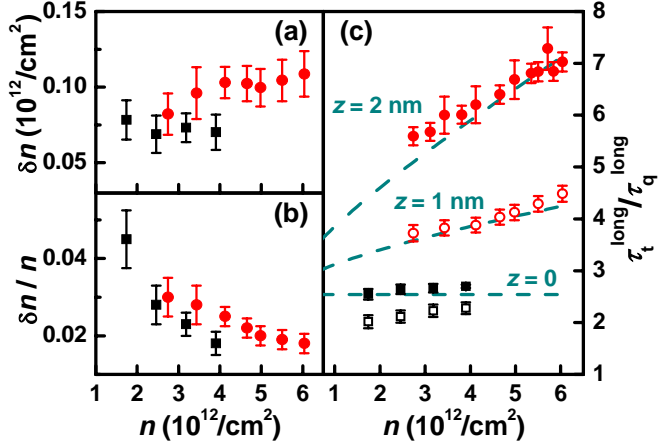

Fig. S3(a) Density inhomogeneity $\delta n$ and (b) $\delta n/n$ vs. $n$ for samples A (squares) and B (circles). (c) $\tau_t^{long}/\tau_q^{long}$ vs. $n$ before (solid symbols) and after correction (open symbols) for samples A (squares) and B (circles). From bottom to top: Dashed lines are calculations for $z = 0$, 1, and 2 nm from Ref. [3].

## 4. Determine the concentration of charged impurity $n_{imp}$ at a distance $z$

We calculate $\tau_t^{long}$ and the corresponding $\sigma_{long}$ by numerically integrating Eq. (3) in Ref. [4] with $z$ extracted from the $\tau_t^{long}/\tau_q^{long}$ ratio. $n_{imp}$ and $\rho_{short}$ are used as fitting parameters to fit the total $\sigma(n)$ using Eq. (2) in the main text. For example, Fig. S4 shows such a fitting to $\sigma(n)$ of sample B with $z = 1$ nm and $n_{imp} = 7.7 \times 10^{11}/cm^2$. Table I lists $n_{imp}$ obtained through this method.

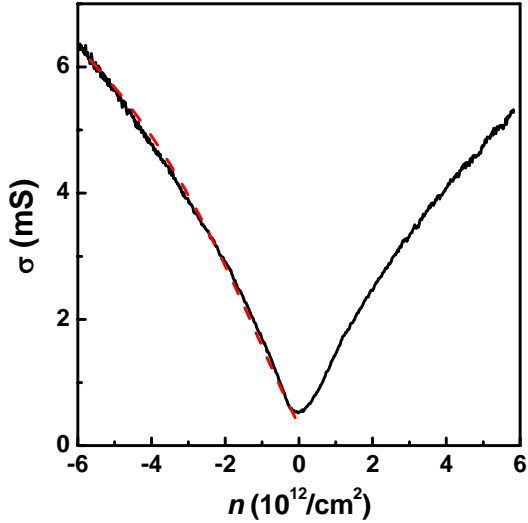

Fig. S4 Solid line: $\sigma(n)$ of sample B. Dashed line: Calculations based on Ref. [3, 4] using $n_{imp} = 7.7 \times 10^{11}/cm^2$, $\rho_{short} = 100\ \Omega$, and $z = 1$ nm.

## 5. Scattering from charges in the bulk of the SiO$_2$ substrate

Although $\tau_t^{long}/\tau_q^{long}$ in our samples can be accounted by a spacer layer of thickness $z$ between graphene and the SiO$_2$ substrate, such data does not rule out contributions from charges residing in the bulk of the SiO$_2$. Since both $\tau_t$ and $\tau_q$ are smoothly varying functions of $z$, the total effect from distributed charges can manifest as an average $z$ with corresponding $\tau_t^{long}/\tau_q^{long}$. To assess the likelihood of this scenario, we have calculated $\tau_t^{long}$, $\tau_q^{long}$, and $\tau_t^{long}/\tau_q^{long}$ caused by charged impurities uniformly distributed within a certain distance $d$ of the graphene sheet, and estimated the effective 2D density of the impurities $n^{eff}_{imp}$ in this layer. Figure S5 plots experimental $\tau_t^{long}/\tau_q^{long}$ and calculations based on the above model for samples B and D, for which we have determined $z = 1$ nm, $n_{imp} = 7.7 \times 10^{11}/cm^2$ (B) and $z = 2$ nm, $n_{imp} = 7 \times 10^{11}/cm^2$ (D) previously assuming a δ-layer of impurity. Our calculations show that the $\tau$ data in sample B can also be described by $n^{eff}_{imp} = 8.5 \times 10^{11}/cm^2$ and $d = 3$ nm and the $\tau$ data in sample D are consistent with $n^{eff}_{imp} = 1.1 \times 10^{12}/cm^2$ and $d = 10$ nm (Fig. S4).

Our $SiO_2/Si$ wafers are obtained from a commercial source, which specifies the density of mobile ions in the oxide to be $< 1 \times 10^{10}/cm^2$. The oxide charges in current MOSFETs are generally in the low $10^{11}/cm^2$ regime, the majority of which are specific to the $Si$-$SiO_2$ interface, not the bulk of the oxide [5]. Although we do not know the density of oxide charges in our $SiO_2$ substrate precisely, the above general argument suggests that they are likely to be in the $10^{10}$-$10^{11}/cm^2$ range and are therefore too small to account for the observed scattering times. In addition, the large variation in the values of $d$ needed to explain our data is difficult to reconcile within the bulk oxide charge scenario.

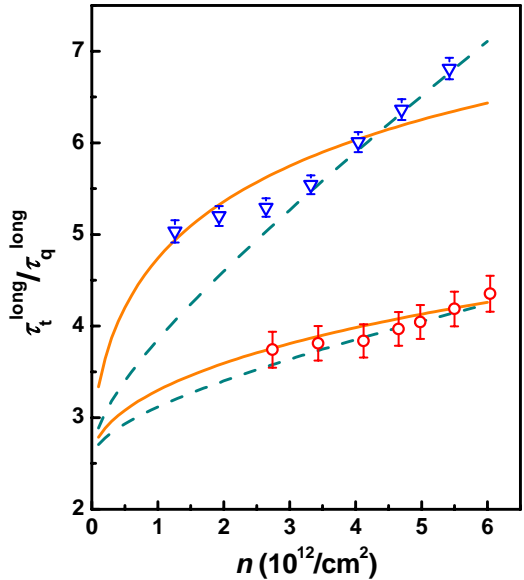

Fig. S5 The measured $\tau_t^{long}/\tau_q^{long}$ ($n$) for samples B (circles) and D (triangles). Dashed lines: Calculations based on a δ-layer of impurities located at $z = 1$ nm (bottom) and 2 nm (top). Solid lines: Calculations based on impurities uniformly distributed within $d = 3$ nm (bottom) and 10 nm (top) of graphene.